\documentclass[11pt]{article}
\usepackage{amssymb}

\newtheorem{Lemma}{Lemma}
\newtheorem{Theorem}{Theorem}
\newtheorem{Proposition}{Proposition}
\newtheorem{Corollary}{Corollary}
\newtheorem{Remark}{Remark}
\newtheorem{Definition}{Definition}

\newenvironment{Proof}
{\begin{trivlist} \item[]{\bf Proof. }}%
{\hspace*{\fill}$\rule{.3\baselineskip}{.35\baselineskip}$\end{trivlist}}

\makeatletter \@addtoreset{equation}{section} \makeatother

\newcommand{\C}{\mathbb{C}}
\def\H{\rm H}
\newcommand{\N}{\mathbb{N}}
\newcommand{\R}{\mathbb{R}}
\newcommand{\Z}{\mathbb{Z}}

\def\wt{\widetilde}
\def\th{\theta}
\def\la{\lambda}
\def\ds{\displaystyle}
\def\ep{\varepsilon}

\newcommand{\iu}{\imath}

\newfam\bifam
\font\tenbi=cmmib10 scaled \magstep1 \font\sevenbi=cmmib10 at 11pt
\font\fivebi=cmmib10 at 6pt \textfont\bifam = \tenbi
\scriptfont\bifam= \sevenbi \scriptscriptfont\bifam= \fivebi

\usepackage[dvips]{epsfig}
\usepackage{graphicx}

\begin{document}
\title{\bf Travelling Waves in Hamiltonian Systems on 2D Lattices with Nearest Neighbor Interactions\footnote{Nonlinearity 2007(accepted)}}
\author{Michal Fe\v ckan$^{\dagger}$ and Vassilis M. Rothos$^{\dagger \dagger}$\\
{\small $^{\dagger}$ Department of Mathematical Analysis and
Numerical Mathematics, Comenius University,}\\
{\small Mlynsk\'a dolina, 842 48 Bratislava, Slovakia,} \\
{\small and Mathematical Institute, Slovak Academy of Sciences,} \\
{\small \v Stef\'anikova 49, 814 73 Bratislava, Slovakia}\\
{\small $^{\dagger \dagger}$ Department of Mathematics, Physics
and Computational Sciences,}\\
{\small Faculty of Technology, Aristotle University of
Thessaloniki, Thessaloniki 54124, Greece}}

\date{\today}
\maketitle

\begin{abstract}
  We study travelling waves on a two--dimensional
  lattice with linear and nonlinear coupling between nearest particles and a periodic nonlinear
  substrate potential. Such a discrete system can model molecules adsorbed on a substrate crystal surface.
We show the existence of both uniform sliding states and periodic
travelling waves as well in a two-dimensional sine-Gordon lattice
equation using topological and variational methods.
\end{abstract}



\section{Introduction}

Nonlinear lattice models are used in many physical applications
\cite{Scot:03}. Spatial discreteness is obviously the standard
situation for applications in solid state physics. One of the
well-known class of discrete models with a broad variety of
applications, consists of the sine-Gordon dynamical lattices, also
known as Frenkel--Kontorova (FK) models, e.g.\ \cite{PeyKru:84}.
For instance, the FK model is a model for the motion of a
dislocation in a crystal, for an adsorbate layer on the surface of
a crystal, for ionic conductors, for glassy materials, for
charge-density-wave transport, for chains of coupled Josephson
junctions, and for sliding friction as well \cite{SE}. Another
important discrete models are either discrete nonlinear
Schr\"odinger systems \cite{Scot:03} used in a variety of
applications ranging from nonlinear optics to the theory of
molecular vibrations, or a discrete model of Fermi, Pasta and Ulam
(FPU) \cite{FP1,FP2} as a prototypical case of nonlinear
mass-spring chains. Systems of discrete lattices are also studied.
For instance, two coupled one-dimensional (1d) nonlinear lattices
are used in \cite{PPF} to represent the Bernal-Fowler filaments in
ice or more complex biological macromolecules in membranes.

Most lattice systems are not integrable. On the other hand, there
is an important class of completely integrable lattice equations,
such as the Toda lattice, the Ablowitz--Ladik lattice, and the
integrable discrete sine-Gordon lattice \cite{DKYF, Scot:03}.
These lattices possess exact soliton solutions showing that
discreteness of the host media does not preclude the propagation
of localized coherent structures. Discrete solitons and more
specifically kink-like topological excitations are ubiquitous
structures that arise in numerous physical applications ranging
from dislocations or ferroelectric domain walls in solids, to
bubbles in DNA, or magnetic chains and Josephson junctions, among
others \cite{r3,FlWil:98}. Hence one of the most important
problems in discrete systems is the existence of travelling waves
and uniform sliding states \cite{Sn,SE}. Of course, the study of
their stability is also crucial.

Recently, particular interest has been devoted to the dynamics of
structures on two--dimensional (2d) lattice systems. For instance
\cite{CW, FBLH}, coherent localized and extended defects such as
dislocations, domains walls, vortices, grain boundaries, etc.,
play an important role in the dynamical properties of materials
with  applications to the problem of adsorbates deposited on
crystal surfaces, in superlattices of ultrathin layers, or in
large-area Josephson junctions, which are described as 2d FK type
models. On the other hand, a 2d cubic Hamiltonian lattice appears
in elastostatic investigation \cite{FK} modeling a particles
interaction via interatomic potentials, which is a natural 1d
analogy of the 1d FPU lattice.

In this paper, we focus on undamped 2d FK models and their
generalizations. Motivated by \cite{TRP}, we consider an isotropic
two-dimensional planar model where rigid molecules rotate in the
plane of a square lattice. At site $(n,m)$ the
angle of rotation is $u_{n,m}$, each molecule interacts linearly
with its first nearest neighbors and with a nonlinear periodic
substrate potential. Here, ${\rm G}$ is the linear coupling
coefficient and ${\omega}_{0}^{2}$ is the strength of the
potential barrier or square of the frequency of small oscillations
in the bottom of the potential wells. The equation of motion of
the rotator at site $(n,m)$ is
\begin{equation}
\label{2dSG} {\ddot u}_{n,m}={\rm G}[u_{n+1,
m}+u_{n-1,m}-2u_{n,m}+u_{n, m+1}+u_{n,
m-1}-2u_{n,m}]+{\omega}^{2}_{0}{\sin}u_{n,m}
\end{equation}
J.M.Tamga {\it et al.} \cite{TRP} studied how a weak initial
uniform perturbation can evolve spontaneously into nonlinear
localized modes with large amplitudes and investigated the
solitary-wave and particle-like properties of these robust
nonlinear entities.

The purpose of this paper is to show the existence of {\em uniform
sliding states} \cite{SE} and {\em periodic travelling wave
solutions} for 2d discrete models like (\ref{2dSG}) by using
topological and variational methods as well. Variational methods
are used in \cite{A11,FW,P1,S1} for 1d discrete FPU type lattice
equations, while topological methods are applied in \cite{K1} to
1d damped discrete sine-Gordon lattice equations. Center manifold
reduction methods are used in \cite{r6,I,IK} to find travelling
waves on nonlinear lattice equations. Dissipative lattice systems
are investigated in \cite{CMS}. The existence and stability of
solitary waves are studied also in \cite{FP1,FP2} for FPU
lattices.

Our paper has the following structure: Section 2 discusses the
mathematical formulation of the travelling wave solutions in
two-dimensional lattices. The proof of uniform sliding states in
equations like  (\ref{2dSG}) is given in Section 3. There we also
discuss their stability. In Section 4, using variational methods,
we prove the existence of several types of periodic travelling
wave solutions in equations like (\ref{2dSG}). Section 5 is
devoted to multiplicity results of periodic travelling waves.
Applications of our methods to more general nonlinear lattices are
discussed in Section 6.

\section{Formulation of the problem}

In this paper, we consider the infinite system of ODEs
\begin{equation}
\label{1.1} {\ddot u}_{n,m}=({\Delta}u)_{n,m}-f(u_{n,m}),\quad
(n,m)\in {\Z}^{2}
\end{equation}
on the two dimensional integer lattice ${\Z}^{2}$ under the
following conditions for $f\in C^1(\R,\R)$:
\begin{description}
\item[(H1)] $f$ is odd, i.e $f(-x)=-f(x)$
$\forall x\in \R$.
\item[(H2)] $f$ is $2\pi$-periodic, i.e.
$f(x+2\pi)=f(x)$ $\forall x\in \R$.
\end{description}
$\Delta$ denotes the discrete Laplacian defined as
\begin{equation}
\label{eq1.2} ({\Delta}u)_{n,m}=u_{n+1, m}+u_{n-1, m}+u_{n,
m+1}+u_{n, m-1}-4u_{n,m}
\end{equation}
We set
$$ L:=\max_\R |f'(x)|\, .
$$
So $f$ is globally Lipschitz continuous on $\R$, i.e. $
|f(x)-f(y)|\le L|x-y|$ $\forall x,y\in \R$.

For $f(u)=\omega_0^2 \sin u$ and $G=1$, we get the 2d discrete
sine-Gordon lattice equation (\ref{2dSG}).

There is a wealth of structures that arise in problems on a
two--dimensional lattice, stemming from the fact that one may
consider different directions of motion of a travelling wave.

As a result of the symmetry imposed by the lattice ${\Z}^{2}$, the
existence and speed of a wave generally will depend on the
direction ${\rm e}^{\iu\theta}$ of motion, with a special role of
those directions for which the slope $\tan \theta$ is rational.
Let $\theta\in\R$ be given, consider a solution of (\ref{1.1}) of
the form
\begin{equation}
 u_{n,m}(t)=U(n\cos \theta+m\sin \theta -\nu t) \label{eq1.3}
\end{equation}
for some $\nu\in\R$ and $U:\R\to\R$. We may consider solutions of
(\ref{eq1.3}) to be travelling waves on the lattice ${\Z}^{2}$, in
the direction ${\rm e}^{\iu\theta}$. Substitution of (\ref{eq1.3})
into (\ref{1.1}) leads to the equation
\begin{equation}
\nu^2U''(z)=U(z+\cos \theta )+U(z-\cos \theta )+U(z+\sin \theta
)+U(z-\sin \theta ) -4U(z)-f(U(z)) \label {2.1}
\end{equation}
with $z=n\cos \theta+m\sin \theta -\nu t$.

If $\nu\neq 0$, equation (\ref{2.1}) is a functional differential
equation of a mixed type, where "mixed" refers to the fact that
the equation involves both forward and backward translations of
the argument of the solution $U$.

Note, that if $\theta$ is an integer multiple of $\pi/2$, then
(\ref{2.1}) reduces to the equation \cite{r6}
\begin{equation}
\label{eq1.5} \ds\nu^2U''(z)=U(z+1)-2U(z)+U(z-1)-f(U(z))
\end{equation}
On the other hand, if ${\theta}-{\pi}/4$ is an integer multiple of
$\pi/2$, then under the change of variables $\psi(z)=U(2^{-1/2}z)$
equation (\ref{2.1}) is equivalent to the equation
\begin{equation}
\label{eq1.7}
\ds \nu^2{\psi}''(z)={\psi}(z+1)-2{\psi}(z)+{\psi}(z-1)-2^{-1}f({\psi}(z))
\end{equation}
which is the same as (\ref{eq1.5}), but with a rescaled function
of $f(u)$.

For other values of $\theta$, one does not expect any reduction of
(\ref{2.1}) to a simpler equation.

The limit $\nu\to 0$ in  (\ref{2.1}) results in a singular
perturbation problem. Indeed, when $\nu=0$ then (\ref{2.1}) is in
fact a functional-difference equation, not a
functional-differential equation. In this case, the solution is
defined on the set ${\cal Z}\subset\R$ given by
$${\cal Z}=\big\{\,z=n\cos \theta+m\sin \theta,\, (n, m)\in
{\Z}^{2}\,\big\}
$$
The set $\cal Z$ is either a discrete subset or a countable dense
subset of $\R$, depending on whether the quantity $\tan \theta$ is
rational or irrational. Such functional-difference equations were
studied in the series of papers \cite{A1,A2,A3}. We present a
discussion in this direction in the point 4. of Section 6.

\section{Uniform Sliding States}

In this section, we prove the existence of {\em uniform sliding
states} \cite{SE} in a two-dimensional lattice. In particular, we
are looking for a solution $U(z)$ of (\ref{2.1}) with the
properties
\begin{equation}
U(z+T)=U(z)+2\pi ,\quad U(-z)=-U(z)\label{2.2}
\end{equation}
for some $T>0$. $U(z)$ is called {\em the hull function}
\cite{SE}.

\begin{Theorem}\label{T2.1} Let conditions {\rm (H1)} and {\rm (H2)} hold.
Then equation {\rm (\ref{2.1})} has a solution with {\rm
(\ref{2.2})} for any $\nu >0$ and $T>0$ satisfying the following
condition:
\begin{description}
\item[i)] ether $\nu \ge 1$, or $0<\nu<1$ along with
$$
\frac{\pi}{r_{\nu,j}}k\ne T\quad \forall k\in \N,\quad j=1,2,\cdots, n_\nu\, ,
$$
where $r_{\nu,j}$, $j=1,2,\cdots, n_\nu$ are all positive
solutions of the equation
$$
\nu^2=\frac{\sin^2(r_\nu\sin \theta)+\sin^2(r_\nu\cos \theta)}{r_\nu^2}\, .
$$
\end{description}
Moreover, if $\nu > 1$ and $LT^2<4\pi^2(\nu^2-1)$ then this solution is unique.
\end{Theorem}

\begin{Proof} We split $U(z)$ as follows
$$
U(z)=\frac{2\pi}{T}z+x(z)
$$
with $x(z+T)=x(z)$ and $x(-z)=-x(z)$. So we get
\begin{equation}
\begin{array}{l}
\ds\nu^2x''(z)=x(z+\cos \theta )+x(z-\cos \theta )\\
\ds+x(z+\sin \theta )+x(z-\sin \theta )
-4x(z)-f\Big (\frac{2\pi}{T}z+x(z)\Big )\, .\end{array}\label {2.3}
\end{equation}
On a Hilbert space
$$
X_1:=\Big \{x\in L^2_{loc}(\R,\R) \mid x(z+T)=x(z),\quad x(-z)=-x(z)\Big
\}
$$
we define a linear mapping $L_1 : D(L_1) \subset X_1\to X_1$ as
follows
$$
\ds L_1(x)(z):=-\nu^2x''(z)+x(z+\cos \theta )+x(z-\cos \theta
)+x(z+\sin \theta )+x(z-\sin \theta )-4x(z)$$ where
$D(L_1):=X_1\cap W^{2,2}_{loc}(\R,\R)$. We also define the
nonlinear operator $N_1 : X_1\to X_1$ by
$$ N_1(x)(z):=f\Big (\frac{2\pi}{T}z+x(z)\Big )
$$
The Hilbert space $X_1$ is considered with the inner product
$$
<x,y>:=\int_0^Tx(z)y(z)\, dz
$$
and the corresponding norm is denoted by $\|\cdot \|$. Then (\ref{2.3}) has the form
\begin{equation}
L_1x=N_1(x)\, .\label{2.4}
\end{equation}
We want to show that the mapping $L_1$ is invertible and to derive
an upper bound for the norm of its inverse.

According to Appendix, the spectrum $\sigma(L_1)$ is given by
$$
\sigma(L_1)=\Big \{k^2\nu^2\frac{4\pi^2}{T^2}-4\sin^2\Big (\frac{\pi}{T}k\sin \theta\Big )-
4\sin^2\Big (\frac{\pi}{T}k\cos \theta\Big )\mid k\in \N\Big\}\, .
$$
Let us put
$$
g_\theta(r)=\frac{\sin^2(r\sin \theta)+\sin^2(r\cos \theta)}{r^2}\, .
$$
Then
$$
g_\theta (r)\le \frac{\sin^2(r\sin \th )}{r^2}+\frac{\sin^2(r\cos \th)}{r^2}\le 1\,.
$$
So $g_\theta (r)$ attains its maximum 1 on $[0,\infty)$ at $r=0$.

Hence if $\nu \ge 1$ then $0\notin \sigma (L_1)$.

Since $g_\theta (r)\to 0$ as $r\to \infty$, for $0<\nu <1$, there is a finite number of
positive $r_{\nu,j}$, $j=1,2,\cdots, n_\nu$ solving
$$
\nu^2=g_\theta(r)\, .
$$
If assumption i) holds then again $0\notin \sigma(L_1)$.
Furthermore, there is an orthonormal basis $\{e_k\}_{k\in \R}$ of
the corresponding eigenvectors
$$
e_k:=\left \{\frac{1}{\sqrt{T}}\sin\frac{2\pi k}{T}z\right \}_{k\in \N}\, .
$$
If $0\notin \sigma(L_1)$ then for $x(z)=\sum_{k\in \N}c_ke_k$ we get
$$
L^{-1}_1x=\sum_{k\in \N} \frac{c_k}{h_\theta (\pi k/T)}e_k
$$
for
$$
h_\theta (r):=4\Big (\nu^2r^2-\sin^2(r\sin \theta )-\sin^2(r\cos \theta )\Big )\, .
$$
This shows that the mapping $L^{-1}_1 : X_1\to X_1$ is compact
\cite[p. 13]{Br}. Next, the operator $N_1 : X_1\to X_1$ is bounded
and globally Lipschitz continuous with a constant $L$. Then
(\ref{2.4}) has the form
\begin{equation}
x=L^{-1}_1N_1(x)\, .\label{2.5}
\end{equation}
Applying the Schauder fixed point theorem \cite[p. 22]{Br} to
(\ref{2.5}) we obtain the existence part of Theorem \ref{T2.1}.

Furthermore, for $\nu >1$ we have
$$
\|L^{-1}_1x\|^2  = \sum_{k\in \N}\frac{T^4}{16\pi^4k^4}\frac{c_k^2}{(\nu^2-g_\theta (\pi k/T)))^2}
\le \sum_{k\in \N}\frac{T^4}{16\pi^4}\frac{c_k^2}{(\nu^2-1)^2}=
\frac{T^4}{16\pi^4 (\nu^2-1)^2}\|x\|^2\, .
$$
Hence $\|L_1^{-1}\|\le \frac{T^2}{4\pi^2 (\nu^2-1)}$. If
$\frac{T^2L}{4\pi^2 (\nu^2-1)}<1$ then the Banach fixed point
theorem \cite[p. 24]{ChH} gives the uniqueness result of Theorem
\ref{T2.1}.
\end{Proof}

\begin{Remark}\label{R2.2} The function $\wt U(z):=U(-z)$ for $U(z)$
from Theorem {\rm \ref{T2.1}} satisfies
\begin{equation}
\wt U(z+T)=\wt U(z)-2\pi\, .\label{2.6}
\end{equation}
So we also get a solution of {\rm(\ref{2.1})} satisfying {\rm
(\ref{2.6})}.
\end{Remark}

\begin{figure}
\setlength{\unitlength}{1cm}
\begin{picture}(3,6)
\centerline{
\epsfbox{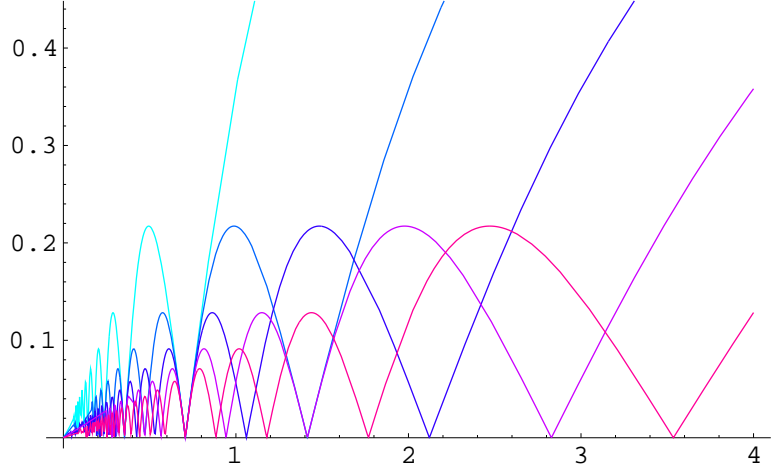}
}
\end{picture}
\caption{The graphs of functions $\wt g_{\pi/4}(\pi k/T)$ for
$k=1,2,3,4,5$ and $0<T<4$.} \label{fig1}
\end{figure}

We note that condition i) is equivalent to $\nu\ne \wt g_\th(\pi
k/T)$ for all $k\in\N$, and a given $T>0$ with
$$
\wt g_\th(r):=\frac{\sqrt{\sin^2(r\sin\th)+\sin^2(r\cos\th)}}{r}\,
.
$$
For a fixed $k\in\N$, the function $T\to \wt g_\th(\pi k/T)$ is
oscillating near $T\sim 0$. So it is impossible to draw a complete
diagram in the $\nu-T$-plane where condition i) holds. We draw
parts of graphs of these functions for $\th=\pi/4$, $k=1,2,3,4,5$
and $0<T<4$ in Figure \ref{fig1}. We can expect from Figure
\ref{fig1} the following result.

\begin{Corollary}\label{C2.1} Let conditions {\rm (H1)} and {\rm
(H2)} hold. If $\nu>T\sqrt{2}/\pi$ then equation {\rm (\ref{2.1})}
has a solution with {\rm (\ref{2.2})}. If
$\nu>T\frac{\sqrt{L+8}}{2\pi}$ then this solution is
unique.\end{Corollary}
\begin{Proof} If $\nu>T\sqrt{2}/\pi$ then from $h_\th(r)\ge 4\nu^2r^2-8$, we get
$\|L_1^{-1}\|\le \frac{T^2}{4\nu^2\pi^2-8T^2}$. This gives
the existence part applying the Schauder fixed point theorem to
(\ref{2.5}). Next, $\nu>T\frac{\sqrt{L+8}}{2\pi}$ gives
$\|L_1^{-1}\|L\le \frac{LT^2}{4\nu^2\pi^2-8T^2}<1$. The use
of the Banach fixed point theorem to (\ref{2.5}) implies the
uniqueness part. We note that the conditions of Corollary \ref{C2.1}
implies condition i) of Theorem \ref{T2.1}.\end{Proof}

Now we improve Theorem \ref{T2.1} as follows:

\begin{Proposition}\label{P2.3} Let conditions {\rm(H1),
(H2)} and {\rm i)} hold. Then equation {\rm (\ref{2.1})} has a
solution $U(z)$ satisfying {\rm(\ref{2.2})} which is accumulated
by similar ones, i.e. there are sequences $\{U_n(z)\}_{n\in\N}$
and $\{T_n\}_{n\in\N}$ such that $U_n(z)$ is a solution of {\rm
(\ref{2.1})} satisfying {\rm(\ref{2.2})} with $T_n$ along with
$T_n\to T$ and $U_n(z)\rightrightarrows U(z)$ uniformly on
$[-2T,2T]$ as $n\to\infty$.\end{Proposition}
\begin{Proof} We know from the proof of Theorem \ref{T2.1} that all
solutions of (\ref{2.4}) belong into a ball $B_0\subset X_1$. If
there are infinitely many solutions of (\ref{2.4}) then their
accumulation point $x(z)$ gives $U(z)=\frac{2\pi}{T}z+x(z)$ which
satisfies properties of Proposition \ref{P2.3}. Here we use:
$L^{-1}_1 : X_1\to D(L_1)\subset W^{2,2}_{loc}(\R,\R)\subset
C(\R,\R)$. If there are finitely many solutions
$\{x_k\}_{k=1}^{M_0}\subset B_0$ of (\ref{2.4}) then from the fact
that the Leray-Schauder topological degree
$\deg(I-L_1^{-1}N_1,B_0,0)=1$ (cf. \cite[p. 57]{Br}, \cite[p.
65]{ChH}), there is an $x_{n_0}$ such that its local
Leray-Schauder topological degree
$\deg(I-L_1^{-1}N_1,x_{n_0},0)\ne 0$ (cf. \cite[p. 69]{ChH},
\cite[p. 192]{M1}). Next, slightly changing $T$, condition i) is
still satisfied and $I-L_1^{-1}N_1$ depends continuously on $T$.
This implies, that for any $\hat T\sim T$, there is a solution
$\hat U(z)=\frac{2\pi}{\hat T}z+\hat x(z)$ of (\ref{2.1})
satisfying (\ref{2.2}) with $\hat T$ for $\hat x(z)$ near
$x_{n_0}(z)$. Consequently, $U(z)=\frac{2\pi}{T}z+x_{n_0}(z)$
satisfies the properties of Proposition \ref{P2.3}. The proof is
finished. \end{Proof}

Next we note that for $\th=0$ we get (\ref{eq1.5}) and
$g_0(r)=\frac{\sin^2r}{r^2}$. Condition i) of Theorem {\rm
\ref{T2.1}} for $0<v<1$ and $T=2\pi$ becomes
$$
\nu \ne \frac{2|\sin k/2|}{k},\quad \forall k\in \N\, .
$$
In terms of \cite[p. 1604]{SE}, this means that condition i) of
Theorem {\rm \ref{T2.1}} is \emph{a superharmonic nonresonance
assumption}. In \cite{SE}, hull functions were calculated
numerically in the 1d FK model with a driving force and with
damping. Theorem {\rm \ref{T2.1}} predicts that these solutions
persist as exact solutions in the absence of a driving force and
damping and in 2d. Moreover, Proposition \ref{P2.3} asserts that
under a superharmonic nonresonance there are continuum many
uniform sliding states accumulated by sequences of uniform sliding
states. Hence they are clearly neither asymptotically stable nor
hyperbolic. To illustrate that how the stability of uniform
sliding states is sophisticated, we consider
\begin{equation}
\nu^2U''(z)=2U\left (z+\frac{\sqrt{2}}{2}\right )+2U\left
(z-\frac{\sqrt{2}}{2}\right)-4U(z)-\ep \sin U(z)\label {4.13}
\end{equation}
with $\ep \ne 0$ small. Taking $T=\sqrt{2}$, $1\ge |\ep|>0$ and
$\nu>\frac{3\sqrt{2}}{2\pi}\doteq 0.6752$, Corollary \ref{C2.1}
ensures the existence of a unique odd $\sqrt{2}$-periodic solution
of
\begin{equation}
\nu^2x''(z)=2x\left (z+\frac{\sqrt{2}}{2}\right )+2x\left
(z-\frac{\sqrt{2}}{2}\right)-4x(z)-\ep \sin \left (\sqrt{2}\pi
z+x(z)\right )\label {2.7}
\end{equation}
with $x(z)=O(\ep)$. Then
$$
u^0_{n,m}(t):=U\left(\frac{\sqrt{2}}{2}(n+m)-\nu t\right)=\pi
(n+m)-\sqrt{2}\pi\nu t+x\left(\frac{\sqrt{2}}{2}(n+m)-\nu t\right)
$$
satisfies $u^0_{n+2,m}(t)=u^0_{n,m+2}(t)=u^0_{n,m}(t)+2\pi$. Hence
we consider (\ref{1.1}) under periodic conditions
\begin{equation}
u_{n+2,m}(t)=u_{n,m+2}(t)=u_{n,m}(t)+2\pi\, .\label {2.8}
\end{equation}
Next we compute
$$
\cos(u^0_{n,m}(t))=\cos(\pi (n+m)-\sqrt{2}\pi\nu
t)+O(\ep)=\cos\pi(n+m)\cos\sqrt{2}\pi\nu t+O(\ep)\, .
$$
Consequently, the linearization of (\ref{1.1}) under conditions
(\ref{2.8}) at $u^0_{n,m}(t)$ up to $O(\ep^2)$-terms has the form
\begin{equation}\label{2.9}
\begin{array}{rl}
\ds \ddot v_{0,0} & =
2v_{1,0}+2v_{0,1}-4v_{0,0}-\ep\cos\sqrt{2}\pi\nu tv_{0,0}\\
\ds \ddot v_{1,0} & = 2v_{0,0}+2v_{1,1}-4v_{1,0}+\ep\cos\sqrt{2}\pi\nu tv_{1,0}\\
\ds \ddot v_{0,1} &
=2v_{0,0}+2v_{1,1}-4v_{0,1}+\ep\cos\sqrt{2}\pi\nu
tv_{0,1}\\
\ds \ddot v_{1,1} & =
2v_{0,1}+2v_{1,0}-4v_{1,1}-\ep\cos\sqrt{2}\pi\nu
tv_{1,1}\,.\end{array}
\end{equation}
For $w:=v_{0,0}-v_{1,1}$, (\ref{2.9}) gives the Mathieu equation
\begin{equation}\label{2.10}
\ddot w+\left(4+\ep\cos\sqrt{2}\pi\nu t\right)w=0\, .
\end{equation}
For $\ep >0$ small, we get a resonance in (\ref{2.10}) at
$\sqrt{2}\pi\nu=4/n$, $n\in\N$, \cite{Ce}. This resonance is
different from superhamonic one. Then from
$\nu>\frac{3\sqrt{2}}{2\pi}$ we obtain $n=1$, i.e. (\ref{2.10}) is
resonant only at $\nu=\nu_0:=2\sqrt{2}/\pi\doteq 0.9003>0.6752$.
Then according to \cite{Ce}, (\ref{2.10}) is stable for $\nu \sim
\nu_0$ in a narrow interval ${\cal I}$, and it is unstable for
$\nu$ outside of ${\cal I}$. The smaller $\ep >0$ the narrow
${\cal I}$. So then (\ref{1.1}) is also unstable at $u^0_{n,m}(t)$
for $\nu$ outside of ${\cal I}$. Of course, for a general $\th$,
the analysis is much more difficult. So we do not carry out it in
this paper.

Finally, sliding states of 1d FK models are investigated also in
\cite{FF, Sn}.

\section{Periodic Travelling Waves}

First, in this Section, we are looking for a solution of (\ref{2.1}) with the property
\begin{equation}
U(z+T)=-U(z)+2\pi\label{3.1}
\end{equation}
for some $T>0$. Of course then
\begin{equation}
U(z+2T)=U(z)\, .\label{3.2}
\end{equation}
We note that conditions (H1) and (H2) imply
\begin{equation}
f(\pi-x)=-f(\pi+x)\, .\label{3.3}
\end{equation}
Hence $f(\pi)=0$ and $x(z)=\pi$ is a trivial/constant solution of (\ref{2.1}) satisfying (\ref{3.1}).

We need the following definition and Theorem 2.8 from \cite{F1}
(see also \cite[p. 185, and p. 216, Exercise 3.]{M1}).

\begin{Definition}\label{D1} Let $\H$ be a Hilbert space and $L : {\H}\to {\H}$ be a symmetric bounded
linear operator. If the dimension of the subspace of all
eigenvectors of $L$  with negative eigenvalues is finite then it
is called as the Morse index of $L$ and it is denoted by ${\rm
index}\, L$.\end{Definition}

\begin{Theorem}\label{P1} Let $<\cdot,\cdot >$ be an inner product on a Hilbert space $\H$
with the corresponding norm $\|\cdot \|$. Let us consider a
function $f\in C^2(\H,\R)$ of a form
$$
f(u)=\frac{1}{2}<Lu,u>+g(u)
$$
with the following properties:
\begin{description}
\item[I)] $L$ is an isomorphism possessing the Morse index, i.e. the dimension of the subspace of all
eigenvectors of $L$  with negative eigenvalues is finite.
\item[II)] $\nabla g : \H\to \H$ is a compact operator.
\item[III)] There is a constant $M>0$ such that $\|\nabla
g(u)\|\le M$ $\forall u\in \H$. \item[IV)] $\nabla f(0)=0$ and the
Morse index of ${\rm Hess}\, f(0)$ exists, i.e. the dimension of the subspace of all
eigenvectors of ${\rm Hess}\, f(0)$  with negative eigenvalues is finite.
\end{description}
If the following condition is satisfied:
$$
{\rm index}\, L\notin [{\rm index \, Hess}\, f(0),{\rm index \, Hess}\, f(0)+\dim \ker {\rm Hess}\, f(0)]
$$
then $f$ possesses a nonzero critical point on $\H$.
\end{Theorem}

\begin{Remark}\label{R3.1} When $L=\alpha I+B$ for a constant $\alpha>0$
and a compact linear symmetric operator $B : H\to H$, then the
both linear operators $L$ and ${\rm Hess}\, f(0)$ have Morse
indices, i.e. the dimensions of the subspaces of all eigenvectors
of $L$ and ${\rm Hess}\, f(0)$  with negative eigenvalues are
finite, respectively (cf. {\rm \cite[p. 75]{Br}}).\end{Remark}

We denote by $(a,b)\subset \R$ the interval with end points $a,b$
for either $a<b$ or $b<a$. Now we can prove the following result.

\begin{Theorem}\label{T3.1} Let conditions {\rm (H1), (H2)} hold and $f'(\pi)\ne 0$.
Moreover, suppose
\begin{description}
\item[a)] either $\nu\ge 1$, or $0<\nu <1$ and
$$
\frac{\pi}{2r_{\nu,j}}(2k-1)\ne T\quad \forall k\in \N,\quad j=1,2,\cdots, n_\nu\, ,
$$
where $r_{\nu,j}$, $j=1,2,\cdots, n_\nu$ are defined in Theorem {\rm \ref{T2.1}}.
\end{description}
Furthermore, we assume
\begin{description}
\item[b)] There exists a $k_0\in \N$ such that
$$\begin{array}{l}
\ds \nu^2\in \Big (g_\theta \Big (\frac{\pi}{2T}(2k_0-1)\Big
),g_\theta \Big (\frac{\pi}{2T}(2k_0-1)\Big
)+\frac{T^2}{\pi^2(2k_0-1)^2}f'(\pi)\Big )\, .\end{array}
$$
\end{description}
Then {\rm (\ref{2.1})} has a nonconstant solution satisfying {\rm (\ref{3.1})}.
\end{Theorem}

\begin{Proof} We split $U(z)$ as follows
$$
U(z)=\pi+x(z)
$$
with $x(z+T)=-x(z)$. So we get
\begin{equation}
\begin{array}{l}
\ds\nu^2x''(z)=x(z+\cos \theta )+x(z-\cos \theta )\\
\ds+x(z+\sin \theta )+x(z-\sin \theta )
-4x(z)-f(\pi+x(z))\, .\end{array}\label {3.4}
\end{equation}
We take Hilbert spaces
$$
\begin{array}{l}
\ds X_2:=\Big \{x\in L^2_{loc}(\R,\R) \mid x(z+T)=-x(z)\Big \}\\
\ds Y_2:=\Big \{x\in W^{1,2}_{loc}(\R,\R) \mid x(z+T)=-x(z)\Big \}\end{array}
$$
with the above inner product $<x,y>$ and
$$
(x,y):=\int_0^Tx'(z)y'(z)\, dz\, ,
$$
respectively. Moreover, we consider operators
$$
\begin{array}{l}
\ds L_2x:=x(z+\cos \theta )+x(z-\cos \theta )\\
\ds \qquad \qquad +x(z+\sin \theta )+x(z-\sin \theta )-4x(z)\\
\ds N_2(x):=-f(\pi+x(z))\, .\end{array}
$$
On $Y_2$ we set the function
$$
\ds H(x):=\int^{T}_{0}\Big \{\nu^2\frac{x'(z)^2}{2}+(x(z+\cos
\theta )+x(z+\sin \theta))x(z)-2x(z)^2-F(\pi+x(z))\Big \}\, dz
$$
for $F(x)=\int_0^xf(s)\, ds$. Note $H\in C^2(Y_2,\R)$ and
\begin{eqnarray*}
\ds (DH(x), v)&=&\int_0^T\Big \{\nu^2 x'(z)v'(z)+(x(z+\cos \theta
)+x(z+\sin \theta))v(z)\\ \ds&+&(v(z+\cos \theta )+v(z+\sin
\theta))x(z) -4x(z)v(z)-f(\pi+x(z))v(z)\Big \}\, dz
\end{eqnarray*}
If $x$ is a critical point of $H$, then, for every $v\in Y_2$,
\begin{eqnarray}\label{hes1}
\ds 0&=&(DH(x), v)=\int_0^T\Big \{\nu^2 x'(z)v'(z)+(x(z+\cos
\theta )+x(z+\sin \theta))v(z)\cr \ds &+& (v(z+\cos \theta
)+v(z+\sin \theta))x(z) -4x(z)v(z)-f(\pi+x(z))v(z)\Big \}\, dz
\end{eqnarray}
with $x\in Y_2$. But for any $\rm c$, we have
$$
\ds \int^{T}_{0}x(z+{\rm c})v(z)\,dz=\int^{T+{\rm
c}}_{\rm c}x(s)v(s-{\rm c})\,ds=\int^{T}_{0}x(s)v(s-{\rm c})\,ds
$$
since $x, v\in Y_{2}$, gives the $T$-periodicity of $s\to
x(s)v(s-{\rm c})$. Consequently, (\ref{hes1}) gives us $x\in
Y_2\cap W^{2,2}_{loc}(\R,\R)$ along with
\begin{equation}\label{hes2}
\ds 0=\int_0^T\Big \{-\nu^2 x''(z)+x(z+\cos \theta )+x(z+\sin
\theta)+ x(z-\cos \theta )+x(z-\sin \theta) -4x(z)-f(\pi+x(z))\Big
\}v(z)\, dz
\end{equation}
Since, (\ref{hes2}) holds for any $v\in Y_{2}$, we see that the
critical points of $H$ are solutions of (\ref{3.4}).

Next, the Riesz representation theorem of bounded linear
functionals on Hilbert spaces \cite{Ru} gives a linear mapping $K
: X_2\to Y_2$ defined by the equality
$$
(Kh,y)=<h,y>\quad \forall y\in Y_2\, .
$$
Then
$$
\begin{array}{l}
\ds \nabla H(x)={\nu}^2 x+KL_2x+KN_2(x)\, ,\\
\ds {\rm Hess}\, H(x)=\nu^2I+KL_2+KDN_2(x)\, ,\end{array}
$$
where we consider
$$
L_2,N_2 : Y_2\hookrightarrow X_2\to X_2\, .
$$
Hence
$$
\wt L_2:=KL_2,\quad \wt N_2(x):=KN_2(x)
$$
are compact operators. Moreover, $\wt N_2 : Y_2\to Y_2$ is bounded.

Since by (\ref{3.3}), $f(\pi)=0$, equation (\ref{3.4}) has a trivial solution $x(z)=0$.
We show the existence of a nontrivial/nonzero one by using Theorem \ref{P1}.
The boundedness of $\wt N_2$ gives that $\nabla H(x)$ is asymptotically
linear at infinity with an asymptote $\bar L_2=\nu^2I+\wt L_2$. Next using an orthogonal basis
$$
\left \{\cos \frac{\pi}{T}(2k-1)z,\sin \frac{\pi}{T}(2k-1)z\right \}_{k\in \N}
$$
we see that
$$\begin{array}{l}
\ds \sigma(\bar L_2)=\Big \{\nu^2-g_\theta \Big (\frac{\pi}{2T}(2k-1)\Big )\Big \}_{k\in \N}\, ,\\
\ds \sigma({\rm Hess}\, H(0))=\Big \{\nu^2-g_\theta \Big
(\frac{\pi}{2T}(2k-1)\Big )-\frac{T^2}{\pi^2(2k-1)^2}f'(\pi)
\Big \}_{k\in \N}\, .\end{array}
$$
Assumption a) gives $\ker \bar L_2=\{0\}$. According to Definition
\ref{D1}, the Morse indices of $\bar L_2$ and ${\rm Hess}\, H(0)$
are given by
$$
\begin{array}{l}
\ds {\rm index}\, \bar L_2=\# \Big \{k\in \N \mid \nu^2<g_\theta
\Big (\frac{\pi}{2T}(2k-1)\Big )\Big \}\times 2\, , \\
\ds {\rm index \, Hess}\, H(0)= \\
\ds \qquad \qquad \# \Big \{k\in \N \mid \nu^2< g_\theta \Big
(\frac{\pi}{2T}(2k-1)\Big )+\frac{T^2}{\pi^2(2k-1)^2}f'(\pi)\Big
\}\times 2 \, .\\
\ds \dim \ker {\rm Hess}\, H(0)=\\
\ds \qquad \qquad \# \Big \{k\in \N \mid \nu^2=g_\theta \Big
(\frac{\pi}{2T}(2k-1)\Big )+\frac{T^2}{\pi^2(2k-1)^2}f'(\pi)\Big
\}\times 2 \, .\end{array}
$$
Hence condition b) implies
$$
{\rm index}\, \bar L_2\notin \Big [{\rm index \, Hess}\, H(0),{\rm
index \, Hess}\, H(0)+\dim \ker {\rm Hess}\, H(0)\Big ]\, .
$$
Then Theorem \ref{P1} ensures the existence of a nonzero critical
point of $H$.
\end{Proof}

Similarly we have the next results.

\begin{Theorem}\label{T3.2} Let conditions {\rm (H1), (H2)} hold and $f'(0)\ne 0$.
Moreover, suppose condition {\rm a)} of Theorem {\rm \ref{T3.1}} along with
\begin{description}
\item[c)] There exists a $k_0\in \N$ such that
$$
\ds \nu^2\in \Big (g_\theta \Big (\frac{\pi}{2T}(2k_0-1)\Big
),g_\theta \Big (\frac{\pi}{2T}(2k_0-1)\Big
)+\frac{T^2}{\pi^2(2k_0-1)^2}f'(0)\Big )\, .
$$
\end{description}
Then {\rm (\ref{2.1})} has a nonzero solution satisfying
\begin{equation}
U(z+T)=-U(z)\, .\label{3.5}
\end{equation}
\end{Theorem}

\begin{Proof} We follow the proof of Theorem \ref{T3.1} for (\ref{2.1}) in place of (\ref{3.4}).
So on $Y_2$ we set the function
$$
\bar H (x):=\int_0^T\Big \{\nu^2\frac{x'(z)^2}{2}+(x(z+\cos \theta
)+x(z+\sin \theta ))x(z)-2x(z)^2-F(x(z))\Big \}\, dz
$$
Then,
$$
\begin{array}{l}
\ds \nabla \bar H(x)=\nu^2 x+KL_2x+KN_3(x)\, ,\\
\ds {\rm Hess}\, H(x)=\nu^2I+K_2L_2+KDN_3(x)\, ,\end{array}
$$
for
$$
N_3(x):=-f(x(z))\, .
$$
Consequently, we get
$$
\begin{array}{l}
\ds {\rm index \, Hess}\, \bar H(0) = \# \Big \{k\in \N \mid
\nu^2< g_\theta \Big (\frac{\pi}{2T}(2k-1)\Big
)+\frac{T^2}{\pi^2(2k-1)^2}f'(0)\Big \}\times 2 \, ,\\
\ds \dim \ker {\rm Hess}\, \bar H(0)= \# \Big \{k\in \N \mid \nu^2
= g_\theta \Big (\frac{\pi}{2T}(2k-1)\Big
)+\frac{T^2}{\pi^2(2k-1)^2}f'(0)\Big \}\times 2 \, .\end{array}
$$
Hence condition c) implies
$$
{\rm index}\, \bar L_2\notin \Big [{\rm index \, Hess}\, \bar
H(0),{\rm index \, Hess}\, \bar H(0) + \dim \ker {\rm Hess}\, \bar
H(0)\Big ]\, .
$$
Assumption a) of Theorem \ref{T3.1} gives $\ker \bar L_2=\{0\}$.
Then, Theorem \ref{P1} ensures the existence of a nonzero critical
point of $\bar H$.
\end{Proof}

\begin{Remark}\label{R3.3} Conditions {\rm (\ref{3.1})} and {\rm
(\ref{3.5})} imply condition {\rm (\ref{3.2})}.
But they give two different solutions of {\rm (\ref{2.1})}
satisfying {\rm (\ref{3.2})}.
\end{Remark}

\begin{Theorem}\label{T3.4} Let conditions {\rm (H1), (H2)} hold and $f'(0)\ne 0$.
Moreover, suppose condition {\rm i)} of Theorem {\rm \ref{T2.1}}
along with
\begin{description}
\item[d)] There exists a $k_0\in \N$ such that
$$
\ds \nu^2\in \Big (g_\theta \Big (\frac{\pi}{T}k_0\Big ),g_\theta
\Big (\frac{\pi}{T}k_0\Big )+\frac{T^2}{4\pi^2k_0^2}f'(0)\Big )\,
.
$$
\end{description}
Then {\rm (\ref{2.1})} has a nonzero solution satisfying
\begin{equation}
U(z+T)=U(z),\quad U(-z)=-U(z)\, .\label{3.7}
\end{equation}
\end{Theorem}

\begin{Proof} We consider (\ref{2.1}) on the Hilbert spaces $X_1$ and
$$
Y_1:=\Big \{x\in W^{1,2}_{loc}(\R,\R) \mid x(z+T)=x(z),\quad x(-z)=-x(z)\Big
\}\, ,
$$
where $Y_1$ is endowed with the above inner product $(\cdot ,\cdot )$.

Next, on $Y_1$ we take the above function $\bar H$. The rest of
the proof is a modification of Theorem \ref{T3.1}, but we present
it here for completeness. Again, the Riesz representation theorem
gives a linear mapping $\bar K : X_1\to Y_1$ defined by the
equality
$$
(\bar Kh,y)=<h,y>\quad \forall y\in Y_1\, .
$$
So
$$
\begin{array}{l}
\ds \nabla \bar H(x)=\nu^2 x+\bar KL_2x+\bar KN_3(x)\, ,\\
\ds {\rm Hess}\, \bar H(x)=\nu^2I+\bar KL_2+\bar KDN_3(x)\, ,\end{array}
$$
where we consider
$$
L_2,N_3 : Y_1\hookrightarrow X_1\to X_1\, .
$$
Clearly
$$
\hat L_2:=\bar KL_2,\quad \hat N_3(x):=\bar KN_3(x)
$$
are compact operators. Moreover, $\nabla \bar H(x)$ is
asymptotically linear at infinity with an asymptote $\breve
L_2=\nu^2I+\hat L_2$. Using the orthogonal basis $\{e_k\}_{k\in
\N}$, we get
$$
\begin{array}{l}
\ds {\rm index}\, \breve L_2=\# \Big \{k\in \N \mid \nu^2<g_\theta
\Big (\frac{\pi}{T}k\Big )\Big \}\, ,\\
\ds {\rm index \, Hess}\, \bar H(0) = \# \Big \{k\in \N \mid \nu^2
  < g_\theta \Big (\frac{\pi}{T}k\Big )+
\frac{T^2}{4\pi^2k^2}f'(0)\Big \} \, ,\\
\ds \dim \ker {\rm Hess}\, \bar H(0)= \# \Big \{k\in \N \mid \nu^2
= g_\theta \Big (\frac{\pi}{T}k\Big )+
\frac{T^2}{4\pi^2k^2}f'(0)\Big \} \, .\end{array}
$$
Hence condition d) implies
$$
{\rm index}\, \breve L_2\notin \Big [{\rm index \, Hess}\, \bar
H(0),{\rm index \, Hess}\, \bar H(0) + \dim \ker {\rm Hess}\, \bar
H(0)\Big ]\, .
$$
Assumption i) of Theorem \ref{T2.1} gives $\ker \breve L_2=\{0\}$.
Then Theorem \ref{P1} ensures the existence of a nonzero critical
point of $\bar H$.
\end{Proof}

Applying the above results, for instance, we have the following

\begin{Corollary}\label{C1} Let conditions {\rm (H1)} and {\rm (H2)} hold. If $\nu\ge 1$ and
$$
f'(0)>\frac{\pi^2}{T^2}\Big (\nu^2-g_\theta \Big
(\frac{\pi}{2T}\Big )\Big )
$$
then {\rm (\ref{2.1})} has a nonzero solution satisfying {\rm
(\ref{3.5})}.\end{Corollary}

\begin{Corollary}\label{C2} Let conditions {\rm (H1)} and {\rm (H2)}
hold. If $\nu\ge 1$ and
$$
f'(0)>\frac{4\pi^2}{T^2}\Big (\nu^2-g_\theta \Big
(\frac{\pi}{T}\Big )\Big )
$$
then {\rm(\ref{2.1})} has a nonzero odd $T$-periodic
solution.\end{Corollary}

Now let $\theta =\pi/4$. Then there is an $\nu_1\in (0,1)$ such
that for any $\nu_1<\nu<1$, the equation
$$
\nu^2=g_{\pi/4}(r)=2\frac{\sin^2\left(\frac{\sqrt{2}}{2}r\right)}{r^2}
$$
has a unique positive solution $r_\nu$. An approximate value of
$\nu_1$ is 0.2172, and $r_\nu\in (0,3.6112)$. Taking step by step
$k_0=2,1,1$ in Theorems \ref{T3.1}, \ref{T3.2} and \ref{T3.4}, we
can derive from these theorems the following result.

\begin{Corollary}\label{T3.5} Let $\theta=\pi/4$ and $\nu\in (\nu_1,1)$. If
\begin{equation}
\begin{array}{l}
\ds \frac{\pi}{r_\nu}<T<\frac{3\pi}{2r_\nu}\, ,\\
\ds f'(\pi )>\Big (\nu^2-g_{\pi/4}\Big (\frac{3\pi}{2T}\Big )\Big )\frac{9\pi^2}{T^2}\, ,\\
\ds f'(0)<\frac{\pi^2}{T^2}\min \Big \{\nu^2-g_{\pi/4}\Big
(\frac{\pi}{2T}\Big ),4\Big (\nu^2-g_{\pi/4}\Big
(\frac{\pi}{T}\Big )\Big )\Big \}\,
.\end{array}\label{3.8}
\end{equation}
Then {\rm (\ref {2.1})} has $4$ nontrivial/nonconstant solutions
satisfying either {\rm (\ref {2.2})}, or {\rm (\ref {3.1})}, or
{\rm (\ref {3.5})}, or {\rm (\ref {3.7})},
respectively.\end{Corollary}

On the other hand, taking $k_0=2$ in Theorem \ref{T3.2} we get

\begin{Corollary}\label{T3.51} Let $\theta=\pi/4$ and $\nu\in (\nu_1,1)$. If
\begin{equation}
\begin{array}{l}
\ds \frac{\pi}{r_\nu}<T<\frac{3\pi}{2r_\nu}\, ,\\
\ds f'(0)>\Big (\nu^2-g_{\pi/4}\Big (\frac{3\pi}{2T}\Big )\Big
)\frac{9\pi^2}{T^2}\, .\end{array}\label{3.81}
\end{equation}
Then {\rm (\ref {2.1})} has a nonzero solution satisfying {\rm
(\ref {3.5})}.\end{Corollary}

To be more concrete, we take $\theta =\pi/4$ and $\nu=1/2$. Then
we can numerically verify that approximately $r_{1/2}\doteq
2.6806$ and then the conditions of (\ref{3.8}) hold for
$$
1.17196<T<1.7579,\quad f'(0)<-5,\quad f'(\pi)>16\, ,
$$
while the conditions of (\ref{3.81}) hold for
$$
1.17196<T<1.7579,\quad f'(0)>16\, .
$$
So we obtain the following result.

\begin{Corollary}\label{T3.6} For any $\omega >16$ and $1.17196<T<1.7579$, the $2$d discrete sine-Gordon equation
\begin{equation}
\ddot
u_{n,m}-u_{n+1,m}-u_{n-1,m}-u_{n,m+1}-u_{n,m-1}+4u_{n,m}+\omega
\sin u_{n,m}=0\label{3.9}
\end{equation}
possesses $4$ nontrivial/nonconstant travelling wave solutions of
the form
$$
u_{n,m}(t)=\pi+U\Big (\frac{1}{\sqrt{2}}\Big (n+m\Big
)-\frac{1}{2}t\Big )
$$
for $U(z)$ satisfying either {\rm (\ref {2.2})}, or {\rm (\ref
{3.1})}, or {\rm (\ref {3.5})}, or {\rm (\ref {3.7})},
respectively.
\end{Corollary}
\begin{Proof} First we change $u=w+\pi$. Then (\ref{3.9})
becomes
\begin{equation}
\ddot
w_{n,m}-w_{n+1,m}-w_{n-1,m}-w_{n,m+1}-w_{n,m-1}+4w_{n,m}-\omega
\sin w_{n,m}=0\, .\label{3.10}
\end{equation}
Now we take $f(w)=-\omega \sin w$. Then $f'(0)=-\omega <0$ and
$f'(\pi)=\omega >0$. The results follow from the above
observation. The proof is finished.\end{Proof}

We note that the solution $\bar U(z):=\pi +U(z)$ of (\ref{3.9})
corresponding to condition (\ref{3.1}) in Corollary \ref{T3.6}
satisfies $\bar U(z+T)=-\bar U(z)+4\pi$. So $\bar
U(z+T)-2\pi=-\Big (\bar U(z)-2\pi)\Big )$, and $\wt V(z)=\bar
U(z)-2\pi$ determines a $2T$-periodic travelling wave for
(\ref{3.9}) satisfying (\ref{3.5}).

Of course, more layers of continuum many uniform sliding states
and periodic travelling waves can be shown than in Corollaries
\ref{T3.5}, \ref{T3.51} and \ref{T3.6} by choosing different $k_0$
and $\th$ in Theorems \ref{T3.1}, \ref{T3.2} and \ref{T3.4}.

\section{Multiplicity Results}

It is not difficult to see from \cite{M2}, that $U(z)=0$ is the
only odd $T$-periodic solution of (\ref{2.1}) when $(-L,L)\cap
\sigma(L_1)=\emptyset$. On the other, we do not know in general
about uniqueness or multiplicity of nonconstant/nonzero solutions
in Theorems \ref{T3.1}, \ref{T3.2} and \ref{T3.4}. But it seems
that enforcing assumptions, we could get rather multiplicity
results of nonconstant/nonzero solutions than uniqueness. As an
illustration, now we extend Theorem \ref{T3.4} as follows.

\begin{Theorem}\label{T41.1} Let assumptions {\rm (H1), (H2)} hold, $\nu>1$ and $f'(0)>0$. Moreover, suppose
\begin{description}
\item[e1)] $\nu^2\ne g_\theta
\Big (\frac{\pi}{T}k\Big )+\frac{T^2}{4\pi^2k^2}f'(0)$ $\forall
k\in\N$,
\item[e2)] $\# \Big \{k\in \N \mid \nu^2<g_\theta \Big (\frac{\pi}{T}k\Big )+
\frac{T^2}{4\pi^2k^2}f'(0)\Big \}\ge \Big
[\frac{T\sqrt{L}}{2\pi\sqrt{\nu^2-1}}\Big]\ge 2$, where $[\cdot]$ is the integer part function.
\end{description}
Then {\rm (\ref{2.1})} has at least $2$ nonzero odd $T$-periodic
solutions.\end{Theorem}

\begin{Proof} We work in the framework of Theorem \ref{T3.4}. First, clearly Theorem \ref{T3.4} can be  applied. So under the conditions of Theorem \ref{T41.1}, we have an odd nonzero
$T$-periodic solution $x_0(z)$ of {\rm (\ref{2.1})}. In order to
show another one, we follow the proof of \cite[Theorem 9.2]{M1},
i.e. we must prove
\begin{equation}\label{41.2}
|{\rm index}\, \breve L_2 - {\rm index \, Hess}\, \bar H(0)| \ge
  \dim \ker {\rm Hess}\, \bar H(x_0)\ge 2\, .
\end{equation}
Since $\nu>1$, ${\rm index}\, \breve L_2=0$. Next,
e1) implies $\dim \ker {\rm Hess}\, \bar H(0) = 0$, and we know
$$
{\rm index \, Hess}\, \bar H(0)= \# \Big \{k\in \N \mid \nu^2<
g_\theta \Big (\frac{\pi}{T}k\Big )+
\frac{T^2}{4\pi^2k^2}f'(0)\Big \}\, .
$$
Next, we note that $\ker {\rm Hess}\, \bar H(x_0)$ is determined by the
equation
\begin{equation}
\nu^2x''(z)=x(z+\cos \theta )+x(z-\cos \theta )+x(z+\sin \theta
)+x(z-\sin \theta ) -4x(z)-f'(x_0(z))x(z)\, .\label {41.3}
\end{equation}
We have $\dim\ker {\rm Hess}\, \bar H(x_0)\le \Big
[\frac{T\sqrt{L}}{2\pi\sqrt{\nu^2-1}}\Big]$ from Lemma \ref{L41.1}
bellow. Hence (\ref{41.2}) follows from Lemma \ref{L41.1}
and assumption e2). The proof is finished.\end{Proof}

\begin{Lemma}\label{L41.1} The dimension of all odd $T$-periodic solutions of {\rm
(\ref{41.3})} is less or equals to $\Big
[\frac{T\sqrt{L}}{2\pi\sqrt{\nu^2-1}}\Big]$.
\end{Lemma}
\begin{Proof} Let $k_0=\Big
[\frac{T\sqrt{L}}{2\pi\sqrt{\nu^2-1}}\Big]+1$. Split $X_1=X_{k_0}\oplus
X_{k_0}^\perp$ with $X_{k_0}:=\textrm{span}\{e_k\}_{k=1}^{k_0-1}$.
Let $P_{k_0} : X_1\to X_1$ be the orthogonal projection onto
$X_{k_0}$. Next, for $x=x_1+x_2$, $x_1\in X_{k_0}$, $x_2\in
X_{k_0}^\perp$ from (\ref{41.3}) we obtain
\begin{equation}
\begin{array}{rl}
\ds \nu^2x_1''(z)= & x_1(z+\cos \theta )+x_1(z-\cos \theta
)+x_1(z+\sin \theta )+x_1(z-\sin \theta ) -4x_1(z)
\\
\ds & -P_{k_0}f'(x_0(z))(x_1(z)+x_2(z))\, ,\end{array}\label
{41.4}
\end{equation}
\begin{equation}
\begin{array}{rl}
\ds \nu^2x_2''(z)= & x_2(z+\cos \theta )+x_2(z-\cos \theta
)+x_2(z+\sin \theta )+x_2(z-\sin \theta )
-4x_2(z)\\
\ds & -(I-P_{k_0})f'(x_0(z))(x_1(z)+x_2(z))\, . \end{array}\label
{41.5}
\end{equation}
Let
$$
L_{k_0}:=-\nu^2x_2''(z)+x_2(z+\cos \theta )+x_2(z-\cos \theta
)+x_2(z+\sin \theta )+x_2(z-\sin \theta ) -4x_2(z)
$$
with $D(L_{k_0})=X_{k_0}^\perp\cap W^{2,2}_{loc}(\R,\R)$. Now,
like in the proof of Theorem \ref{T2.1} we derive
$$
\|L_{k_0}^{-1}\|\le \frac{T^2}{4\pi^2k_0^2(\nu^2-1)}\, .
$$
Consequently, (\ref{41.5}) has the form
\begin{equation}
x_2-L_{k_0}^{-1}(I-P_{k_0})f'(x_0)x_2=L_{k_0}^{-1}(I-P_{k_0})f'(x_0)x_1\,
.\label {41.6}
\end{equation}
Since
$$
\|L_{k_0}^{-1}(I-P_{k_0})f'(x_0)x_2\|\le \|L_{k_0}^{-1}\|
\|I-P_{k_0}\| \|f'(x_0)x_2\|\le
\frac{T^2L}{4\pi^2k_0^2(\nu^2-1)}\|x_2\|\, ,
$$
and $\frac{T\sqrt{L}}{2\pi k_0\sqrt{\nu^2-1}}<1$, the Banach fixed
point theorem gives a unique solution $x_2=x_2(x_1)$ of
(\ref{41.6}). Summarizing, the solvability of (\ref{41.3}) is
reduced to (\ref{41.4}) with $x_2=x_2(x_1)$. Since $\dim
X_{k_0}=\Big [\frac{T\sqrt{L}}{2\pi\sqrt{\nu^2-1}}\Big]$, the
proof is finished.
\end{Proof}

\begin{figure}
\setlength{\unitlength}{1cm}
\begin{picture}(5,10)
\centerline{ \epsfbox{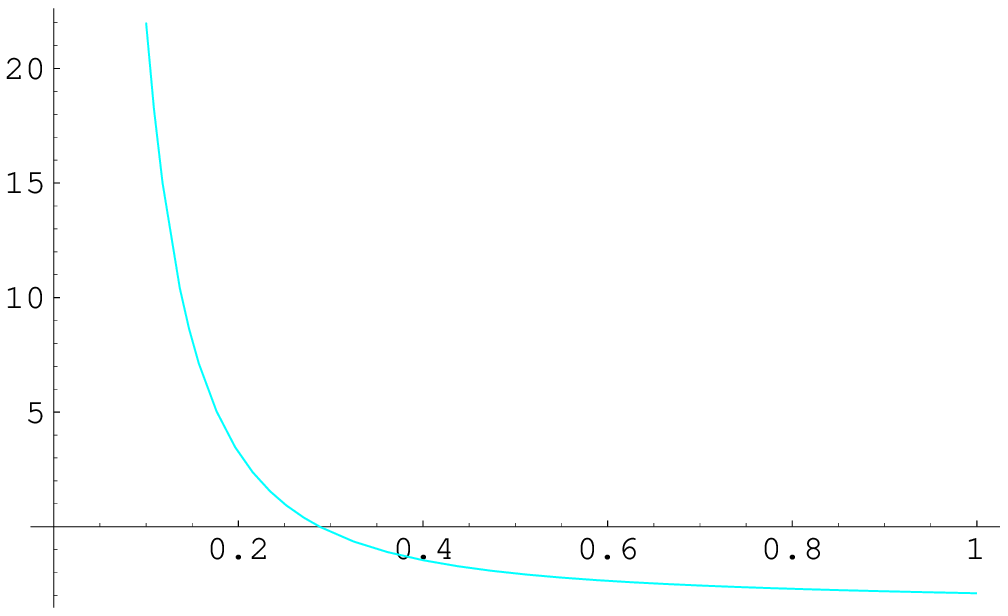} }
\end{picture}
\caption{The graph of (\ref{eee3}).} \label{fig3}
\end{figure}

To show that assumptions e1) and e2) are consistent, we take
$f(x)=\sin x$, so $f'(0)=L=1$. Next, we consider $\theta =\pi/4$,
$\nu=2$ and solve the equation $g(r)=0$ for a function
\begin{equation}\label{eee3}
g(r):=\frac{2\sin^2\Big (r\frac{\sqrt{2}}{2}\Big
)}{r^2}+\frac{1}{4r^2}-4\, ,
\end{equation}
which has a unique solution $r_0$ (see Figure \ref{fig3}) with an
approximative value $0.2880$.

Hence e1) is equivalent to
\begin{equation}\label{cc1}
\frac{\pi}{T}k\ne r_0,\quad \forall k\in \N\, ,
\end{equation}
while e2), when e1) holds, is equivalent to
\begin{equation}\label{cc2}
\Big [\frac{T}{\pi}r_0\Big ]\ge \Big [\frac{T}{\pi}\frac{1}{2\sqrt{3}}\Big ]\ge 2\, .
\end{equation}
We note $\frac{1}{2\sqrt{3}}\doteq 0.2887>r_0$. So the condition
(\ref{cc2}) holds only for bounded $T/\pi$, i.e. $T$ can not be
arbitrarily large, and $\Big [\frac{T}{\pi}r_0\Big ]\le \Big
[\frac{T}{\pi}\frac{1}{2\sqrt{3}}\Big ]$. So it must be $\Big
[\frac{T}{\pi}r_0\Big ]=\Big [\frac{T}{\pi}\frac{1}{2\sqrt{3}}\Big
]$, which gives, according to e1),
$k<\frac{T}{\pi}r_0<\frac{T}{2\sqrt{3}\pi}<k+1$ for a $k\in N$,
$k\ge 2$. That is: $\frac{\pi}{r_0}k<T<2\sqrt{3}\pi(k+1)$. This
implies $\frac{\pi}{r_0}k<2\sqrt{3}\pi(k+1)$, i.e.
$k<\frac{2\sqrt{3}r_0}{1-2\sqrt{3}r_0}\doteq 436.8855$. Hence the
condition (\ref{cc2}) is satisfied on the set
$\bigcup_{k=2}^{436}\left(\frac{\pi}{r_0}k,2\sqrt{3}\pi
(k+1)\right )$ of disjunct intervals. We note that the first
interval with $k=2$ is approximately $(21.8154,32.6484)$ and the
last one with $k=436$ is approximately $(4755.7599,4755.7819)$.
Summarizing we arrive at the following

\begin{Corollary} For $f(x)=\sin x$, $\nu
=2$, $\theta =\pi/4$ and $T\in
\bigcup_{k=2}^{436}\left(\frac{\pi}{r_0}k,2\sqrt{3}\pi (k+1)\right
)$, equation {\rm (\ref{2.1})} has at least $2$ nonzero odd
$T$-periodic solutions.\end{Corollary}

Other multiplicity result can be derived also as follows. We
assume in the above results of Section 4 {\em nonresonance at
infinity}, i.e. conditions i), a) of Theorems \ref{T2.1} and
\ref{T3.1}, respectively. Now we study an opposite case. So let us
again consider (\ref{4.13}) with $\ep \ne 0$ small. We suppose
$\nu_1<\nu<1$ and take
$$
T=\pi/r_\nu\,.
$$
Then (\ref{4.13}) has the form
\begin{equation}
\nu^2x+\hat L_2x+\ep \hat N_3(x)=0,\quad x\in Y_1\, ,\label{4.14}
\end{equation}
in the framework of Theorem \ref{T3.4} with $f(x)=\sin x$ and
$$
\ker (\nu^2I+\hat L_2)=\textrm{im}\, (\nu^2I+\hat L_2)^\perp
=\textrm{span}\, \left \{\sin 2r_\nu z\right \}\, .
$$
Hence (\ref{4.13}) is {\em resonant at infinity, i.e. it is
superharmonic resonant}. Next we split $x(z)=c\sin 2r_\nu
z+x_1(z)$, $x_1\in \textrm{im}\, (\nu^2I+\hat L_2)$ to get from
(\ref{4.14})
\begin{equation}
\nu^2x_1+\hat L_2x_1+\ep P_1\hat N_3\left (c\sin 2r_\nu
z+x_1(z)\right )=0\, ,\label{4.15}
\end{equation}
\begin{equation}
\int_0^{\pi/r_\nu}\sin \left (c\sin 2r_\nu z+x_1(z)\right )\sin
2r_\nu z\, dz=0\, ,\label{4.16}
\end{equation}
where $P_1 : Y_1\to Y_1$ is the orthogonal projection onto
$\textrm{im}\, (\nu^2I+\hat L_2)$. Using the implicit function
theorem \cite[p. 26]{ChH}, we can solve (\ref{4.15}) to get
$x_1(\ep,z)$ for $\ep$ small with $x_1(0,z)=0$. Inserting this
solution into (\ref{4.16}), we get the bifurcation equation
\begin{equation}
Q(\ep,c):=\int_0^{\pi/r_\nu}\sin \left (c\sin 2r_\nu
z+x_1(\ep,z)\right )\sin 2r_\nu z\, dz=0\, .\label{4.17}
\end{equation}
Clearly $Q(0,c)$ is an entire function on the complex plane $\C$
and
$$
Q(0,c):=\frac{\partial}{\partial c}\wt Q(c)
 $$
for
$$
\begin{array}{l}
\ds \wt Q(c):= - \int_0^{\pi/r_\nu}\cos \left (c\sin 2r_\nu
z\right )\, dz=\sum_{n=0}^\infty
\int_0^{\pi/r_\nu}\frac{\sin^{2n}2r_\nu z}{2n!}(-1)^{n+1}\, dz\cdot c^{2n}\\
\ds = \frac{\pi}{r_\nu}\sum_{n=0}^\infty \frac{1.3.5\dots
(2n-1)}{2.4.6\dots
2n}\frac{1}{2n!}(-1)^{n+1}c^{2n}=\frac{\pi}{r_\nu}\sum_{n=0}^\infty
\frac{1}{4^n n!^2}(-1)^{n+1}c^{2n}\, .\end{array}
$$
Hence
$$
\wt Q(\iu \sqrt{c})=-\frac{\pi}{r_\nu}\sum_{n=0}^\infty
\frac{\phi(n)}{n!}c^{n}
$$
for $c\in \C$ and the entire function
$$
\phi(c)=\frac{1}{4^c\Gamma(c+1)}
$$
which has only negative roots $-1,-2,\cdots$. Using arguments of
\cite[Section 8]{Ti}, we see that the entire function
 $\sum_{n=0}^\infty \frac{\phi(n)}{n!}c^{n}$ has only real and
negative roots on $\C$. So $\wt Q(c)$ has only real roots on $\C$
and they are infinitely many. Then again by \cite{Ti}, $Q(0,c)$
has also only real roots on $\C$ separated by roots of $\wt Q(c)$.
This implies that there are infinitely many real roots of $Q(0,c)$
with odd multiplicity. Indeed, if it could exist $c_0$ such that
$Q(0,c)$ would have on $(c_0,\infty)$ only roots with even
multiplicity then $Q(0,c)$ would not change the sign on
$(c_0,\infty)$. So $\wt Q(c)$ would be monotonic on this interval.
This would contradict to that $\wt Q(c)$ has infinitely many roots
on this interval. Here we use the eveness of this function. Next,
each real root of $Q(0,c)$ with odd multiplicity has a nonzero
local Brouwer topological degree \cite[p. 69]{ChH}, which implies
that $Q(\ep,c)$ has infinitely many real roots, so it is
oscillatory on $\R$. Summarizing we arrive at the following

\begin{Theorem}\label{T3.61} Let $\nu_1<\nu<1$. Then
{\rm(\ref{4.13})} has infinitely many odd $\pi/r_\nu$-periodic
solutions $\{U_n(z)\}_{n\in\N}$ with
$$
\left |U_n(z) - c_n\sin 2r_\nu z\right |\le \wt K|\ep|
$$
for $c_n\to \infty$ as $n\to\infty$ and a constant $\wt
K>0$.\end{Theorem}

Theorem \ref{T3.61} states that under superharmonic resonance
$T=\pi/r_\nu$, (\ref{4.13}) possesses infinitely many $T$-periodic
travelling waves. Note $r_\nu\in (0,3.6112)$.

\section{Concluding Remarks}

{\bf 1.}

The same approach as in Section 4, we can apply to
\begin{equation}
\begin{array}{l}
\ds \ddot
u_{n,m}=g(u_{n+1,m}-u_{n,m})-g(u_{n,m}-u_{n-1,m})\\
\ds \qquad \qquad
+g(u_{n,m+1}-u_{n,m})-g(u_{n,m}-u_{n,m-1})-f(u_{n,m})\,
,\end{array}\label {4.1}
\end{equation}
where $f,g\in C^1(\R,\R)$ are bounded odd. Then we study
\begin{equation}
\begin{array}{l}
\ds\nu^2U''(z)=g(U(z+\cos \theta )-U(z))-g(U(z)-U(z-\cos \theta ))\\
\ds+g(U(z+\sin \theta )-U(z))-g(U(z)-U(z-\sin \theta ))-f(U(z))\,
.\end{array}\label {4.2}
\end{equation}

Following the proof of Theorem \ref{T3.4} we have

\begin{Theorem}\label{T4.1} Assuming $f,g\in C^1(\R,\R)$ are
bounded odd along with
\begin{equation}
4\pi^2\nu^2-4\pi^2g'(0)g_\theta\Big (\frac{\pi}{T}\Big
)<T^2f'(0)\, ,\label{4.21}
\end{equation}
then {\rm (\ref{4.2})} possesses a nonzero odd $T$-periodic
solution.\end{Theorem}
\begin{Proof} This result follows like in Theorem \ref{T3.4} for the function
$$
\wt H (x):=\int_0^T\Big \{\nu^2\frac{x'(z)^2}{2}-G(x(z+\cos \theta
)-x(z))-G(x(z+\sin \theta )-x(z))-F(x(z))\Big \}\, dz
$$
with $G(x):=\int_0^xg(s)\, ds$ and $k_0=1$.
\end{Proof}

Moreover, using an approach of \cite{ChH} we obtain

\begin{Theorem}\label{T4.11} Let us assume $f,g\in C^1(\R,\R)$ with $f$
bounded odd, and with $g$ odd satisfying $G(u)\ge c(|u|^{p}-1)$
$\forall u\in \R$ and some constants $p>2$, $c>0$. Moreover we
suppose
\begin{equation}\label{4.223}
\textrm{either $\cos\th/T\notin \Z$ or $\sin\th/T\notin \Z$}
\end{equation}
along with the condition
\begin{equation}\label{4.22}
\min_{k\in\N}\frac{4\pi^2}{T^2}k^2\left(\nu^2-g'(0)g_\th\Big(\frac{\pi
k}{T}\Big)\right)>f'(0)\, .
\end{equation}
Then equation {\rm (\ref{4.2})} possesses a nonzero odd
$T$-periodic solution.\end{Theorem}
\begin{Proof} We consider $\wt H (x)$ over $Y_1$. The condition
(\ref{4.22}) implies $\ker \textrm{Hess}\, \wt H(0)=0$ and
$\textrm{index Hess}\, \wt H(0)=0$, so $0\in Y_1$ is a strict
local minimum of $\wt H$. Next we take
$x_0(z)=\sin\left(\frac{2\pi}{T}z\right)$. Then for any $\alpha>0$
we have
\begin{equation}\label{4.23}
\wt H(\alpha x_0)\le
2\nu^2\alpha^2\frac{\pi^2}{T}-\alpha^{p}\gamma_1+\gamma_2\alpha+\gamma_3
\end{equation}
for
$$
\begin{array}{l}
\ds \gamma_1:=
c\int_0^T\left(\left|\sin\left(\frac{2\pi}{T}(z+\cos\th)\right)-\sin\left(\frac{2\pi}{T}z\right)\right|^{p}+
\left|\sin\left(\frac{2\pi}{T}(z+\sin\th)\right)-\sin\left(\frac{2\pi}{T}z\right)\right|^{p}\right)\,
dz\\
\gamma_2:=T\min_{\R}|f(x)|,\quad \gamma_3:=2cT\, .\end{array}
$$
Note $\gamma_1>0$ according to (\ref{4.223}). Since $\wt H(\alpha
x_0)>0$ for small $\alpha>0$, (\ref{4.23}) implies $\exists
\alpha_0>0$, $\wt H(\alpha_0x_0)=0$. Consequently, the mountain
pass theorem \cite[p. 141]{ChH} can be applied for $\wt H(x)$ to
get its nonzero solution on $Y_1$. The proof is
finished.\end{Proof}

For instance, if \cite{T1}
\begin{equation}\label{nn1}
g(u)=\alpha_1u+\alpha_2u^3,\quad \alpha_{1,2}>0, \quad f(u)=\omega
\sin u\, ,
\end{equation}
where $\alpha_1, \alpha_2$ are the linear and nonlinear coupling
coefficients in the longitudinal and transverse directions,
respectively. Then, (\ref{4.223}) and (\ref{4.22}) are satisfied
when $\tan \th$ is irrational and
$\frac{4\pi^2}{T^2}\left(\nu^2-\alpha_1\right)\ge\omega$. So when
$\omega\le 0$, (\ref{4.2}) with (\ref{nn1}) possesses a nonzero
odd $T$-periodic solution for any $\nu >\sqrt{\alpha_1}$ and any
$T$, $\th$ satisfying (\ref{4.223}). This results can be related
to \cite{P1} and \cite[p. 274, Example (c)]{S1}.

\

{\bf 2.}

Concerning the minimal period of a $T$-periodic solution of
(\ref{2.1}) we derive the following

\begin{Proposition}\label{T4.2}
Let $f$ be globally Lipschitz continuous on $\R$, i.e. $\exists
L>0$, $|f(x)-f(y)|\le L|x-y|$ $\forall x,y\in \R$. Then a period
$T$ of any nonconstant $T$-periodic solution of {\rm (\ref{2.1})}
satisfies
\begin{equation}
T\ge \frac{2\pi \nu}{\sqrt{L+8}}\, .\label{4.3}
\end{equation}
\end{Proposition}
\begin{Proof}Let $P : L^2(0,T)\to L^2(0,T)$ be an orthogonal projection given by
$$
PU:=U(z)-\frac{1}{T}\int_0^TU(s)\, ds\, .
$$
Let us denote by $G_\th(U) : L^2(0,T)\to L^2(0,T)$ the right-hand
side of (\ref{2.1}). We note that $G_\th$ has a global Lipschitz
constant $L+8$. Now we split
$$
U(z)=U_0(z)+c,\quad c\in \R, \quad U_0=PU\, .
$$
We recall that the standard scalar product on $L^2(0,T)$ is
denoted by $<\cdot,\cdot>$ with the corresponding norm
$\|\cdot\|$. Then (\ref{2.1}) implies
\begin{equation}
\begin{array}{l}
\ds \nu^2\frac{4\pi^2}{T^2}\|U_0\|^2\le -\nu^2<U'',U_0>=-<G_\th (U_0+c),U_0>\\
\ds =-<P(G_\th(U_0+c)-G_\th(c)),U_0>\le (L+8)\|U_0\|^2\,
.\end{array}\label{4.4}
\end{equation}
Since $U$ is nonconstant, we get $U_0\ne 0$. Hence (\ref{4.4})
gives (\ref{4.3}).
\end{Proof}

{\bf 3.}

We are able to determine numerical profiles of found travelling
waves only in results where constructive methods are used based on
the Banach fixed point theorem, i.e. in Theorems \ref{T2.1},
\ref{T3.61} and Corollary \ref{C2.1}. We focus here on Theorem
\ref{T2.1} assuming $\nu>1$ and $LT^2<4\pi^2(\nu^2-1)$. Then there
is a unique $x\in X_1$ of (\ref{2.4}). Now like in \cite[p.
270]{SZE}, for any $n\in \N$, we set
$X_{n}:=\textrm{span}\{e_k\}_{k=1}^{n}$. Let $P_{n} : X_1\to X_1$
be the orthogonal projection onto $X_{n}$. Since $L_1P_n=P_nL_1$,
we consider an equation
\begin{equation}
L_1x_n=P_nN_1(x_n),\quad x_n\in X_n\,.\label {4.41}
\end{equation}
Clearly $PN_1 : X_n\to X_n$ has a global Lipschitz constant $L$,
and so $L_1^{-1}PN_1 : X_n\to X_n$ is contractive. Hence
(\ref{4.41}) has a unique solution $x_n\in X_n$. From (\ref{2.4})
and (\ref{4.41}) we derive
$$
L_1(x-x_n)=P_n(N_1(x)-N_1(x_n))+(I-P_n)N_1(x)\, ,
$$
and so
$$
\|x-x_n\|\le L\|L^{-1}_1\| \|x-x_n\|+\|L^{-1}_1(I-P_n)N_1(x)\|\, .
$$
That implies
$$
\|x-x_n\|\le \frac{\|L^{-1}_1(I-P_n)N_1(x)\|}{1-\|L^{-1}_1\|L}\le
\frac{4\pi^2(\nu^2-1)}{4\pi^2(\nu^2-1)-T^2L}\|L^{-1}_1(I-P_n)N_1(x)\|\,
.
$$
Note $\|L^{-1}_1(I-P_n)N_1(x)\|\to\infty$ as $n\to \infty$ and the
convergence rapidity depends  on a smoothness of the function
$z\to N_1(x)(z)=f\left(\frac{2\pi}{T}z+x(z)\right)$.

\begin{figure}

\setlength{\unitlength}{1cm}
\begin{picture}(3,6)
\centerline{
\epsfbox{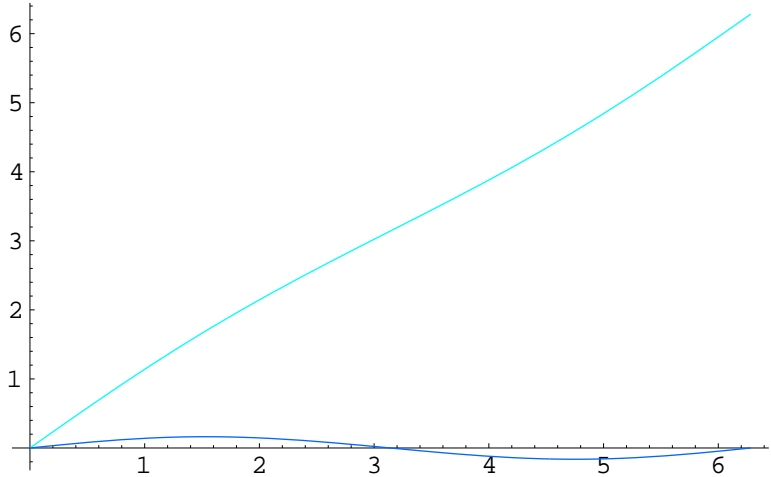}
}
\end{picture}
\caption{Numerical profiles of a uniform sliding state (\ref{2.2})
of (\ref{2.1}) (the mononotone one) and the corresponding odd
$T$-periodic solution of (\ref{2.3}), respectively, for $f(x)=\sin
x$, $T=2\pi$, $\nu=2$, $\th =\pi/4$ and $n=15$ in
(\ref{4.42}).}\label{fig2}
\end{figure}

For $f(x)=\sin x$, $T=2\pi$, $\nu=2$, $\th =\pi/4$ and
$x_n(z)=\frac{1}{\sqrt{2\pi}}\sum_{k=1}^nc_k\sin k z$, equation
(\ref{4.41}) has the form
\begin{equation}\label{4.42}
c_k=\frac{1}{\sqrt{2\pi}\left(4k^2-8\sin^2\frac{\sqrt{2}}{4}k\right
)}\int_0^{2\pi}\sin\left(z+\frac{1}{\sqrt{2\pi}}\sum_{k=1}^nc_k\sin
k z\right )\sin k z\, dz,\quad k=1,\cdots,n\, .
\end{equation}
We know that the right-hand side of (\ref{4.42}) is a contraction,
so it has a unique solution. Numerical profiles for $n=15$ are
drawn on Figure \ref{fig2}.

\

\

{\bf 4.}

For $\nu=0$, equation (\ref{2.1}) becomes a functional-difference
equation for the ground-state of the FK model of the form
\begin{equation}
x(z+\cos \theta )+x(z-\cos \theta )+x(z+\sin \theta )+x(z-\sin
\theta ) -4x(z)=f\Big (\frac{2\pi}{T}z+x(z)\Big )\, .\label {6.61}
\end{equation}
(\ref{6.61}) has complex behavior because it involves small
divisors. This problem was rigorously solved in \cite{A1,A2},
where it was proven that the hull function is monotonous
increasing but it is not necessarily smooth and continuous. It is
smooth only when the period of the ratio $2\pi /T$ is a "good"
irrational number (according to Kolmogorov-Arnold-Moser theory)
and when the potential $f(x)$ is not too large (incommensurate
structures). For larger potentials or for commensurate structures,
the hull function becomes purely discrete. These theorems were
established for the 1d FK model but extend with practically the
same proof in 2d and more dimensions in \cite{A3}, thus including
the 2d model treated in this paper. Thus a natural question
arises: How the uniform sliding states which have continuous hull
functions interpolate with the solutions at zero velocity were the
hull functions may be either continuous or discontinuous? That is
what is a relation between (\ref{2.3}) and (\ref{6.61}) as $\nu\to
0$? We can not answer in general on this question in this paper.
On the other hand, singular delay differential equations are
studied also in \cite{HH}, and we believe that similar approach
could be used to periodic travelling waves as well. More
precisely, we consider (\ref{eq1.5}) and we look for a $(1+\xi
\nu)$-antiperiodic solution of it, i.e. satisfying (\ref{3.5})
with $T=1+\xi \nu$, where $\xi\in \R$ is specified latter and
$\nu$ is small. Then
$$
w(z):=U(-\xi\nu z)
$$
satisfies
\begin{equation}\label{66.1}
\ddot w(z)+\xi^2\left(w(z+1)+2w(z)+w(z-1)+f(w(z))\right)=0 \, .
\end{equation}
For the period of a $\tau$-antiperiodic solution of (\ref{66.1}),
i.e. $w(z+\tau)=-w(z)$, we get the relation
\begin{equation}\label{66.2}
\nu\xi(\tau-1)=1\, .
\end{equation}
Then like in \cite{HH}, we could apply the center manifold
reduction using an approach of \cite{r6,I,IK} near $w(z)\sim 0$ to
derive a reduced ordinary differential equation. The next task
would be to find a $\tau$-antiperiodic solution of the reduced
ordinary differential equation satisfying (\ref{66.2}). We do not
proceed with further computations, but we expect that square waves
could be shown for (\ref{eq1.5}) after embedding $f(u)$ into a
$1$-parametric family $f(u,\lambda)$, $\lambda\in\R$,
$f(u,0)=f(u)$, with certain conditions. The parameter $\xi$ could
be used to determine the dimension of the reduced o.d.e. in a
spectral analysis similar to (\ref{5.5}) in Appendix. So finally,
we could show {\em square travelling waves} for (\ref{1.1}) as
$\nu\to 0$.

 \

{\bf 5.}

We finish this paper with application of our method to a
topological discrete sine-Gordon system \cite{H1} (an 1d lattice) given by
\begin{equation}
\ddot \psi_n=\alpha \cos \psi_n (\sin \psi_{n+1}+\sin\psi_{n-1} )
- \Big (\alpha +\frac{1}{2}\Big )\sin \psi_n(\cos \psi_{n+1}+\cos
\psi_{n-1})\, \label{4.5}
\end{equation}
where $\alpha \in \R$ is a parameter. Inserting $\psi_n(t)=U(n-\nu
t)$ into (\ref{4.5}) we get
\begin{equation}
\nu^2U''(z)=\alpha \cos U(z)(\sin U(z+1)+\sin U(z-1)) - \Big
(\alpha +\frac{1}{2}\Big )\sin U(z)(\cos U(z+1)+\cos U(z-1))\,
.\label{4.6}
\end{equation}

Following the proof of Theorem \ref{T3.4} we have,

\begin{Theorem}\label{T4.3} For any $\nu >0$, $T>0$ and $\alpha \in \R$ satisfying
\begin{equation}
\frac{4\pi^2\nu^2}{T^2}-4\alpha \sin^2\frac{\pi}{T}<1\,
,\label{4.7}
\end{equation}
{\rm (\ref{4.6})} possesses a nonzero odd $T$-periodic
solution.\end{Theorem}
\begin{Proof} Now we take the function
$$
{\breve H}(x):=\int_0^T\Big \{\nu^2\frac{x'(z)^2}{2}+\alpha \sin
U(z+1)\sin U(z) +\Big (\alpha +\frac{1}{2}\Big )\cos U(z)\cos
U(z-1)\Big \}\, dz
$$
and $k_0=1$ in the proof of Theorem \ref{T3.4}.
\end{Proof}

We note \cite{H1} that the "continuum limit" of (\ref{4.6}) is
$\alpha \to -\infty$. Then condition (\ref{4.7}) holds for
$\alpha <0$ and $T>2\pi\sqrt{\nu^2-\alpha}$. On the other hand, the "anti-continuum limit" of
(\ref{4.6}) is $\alpha \to 0$. Then condition (\ref{4.7}) holds
for $|\alpha |<1/4$ and
$$
T>\frac{2\pi \nu}{\sqrt{1-4|\alpha|}}
$$
with $\nu >0$.

\
\section*{Acknowledgments}
This work was supported by the IACM FORTH Greece and the Grant
VEGA-SAV 2/4135/25. We also thanks to the referees for valuable
comments and suggestions which substantially improved our paper.

\appendix

\section{Spectral analysis of certain linear operators}

The spectrums of the above linear self-adjoint bounded operators
can be easily computed, since standard orthogonal basis of the
corresponding functional Hilbert spaces form their eigenvectors.
This follows from the next facts:

\begin{description}
\item [1. -] these orthogonal basis are created by functions of a form $c_1\cos
a_1z+c_2\sin a_1z$ for constants $c_{1,2}$, $a_{1}$, \item[2. -]
these linear operators are linear combinations of the shift
mapping $x(z)\to x(z+a_2)$ and the differential operator $x(z)\to
x''(z)$ for a constant $a_2$, \item[3. -] subspaces $\textrm
{span} \, \{\cos a_1z,\sin a_1z\}$ are invariant under operators
from point 2.
\end{description}

For instance, in Theorem \ref{T2.1}, the standard basis
$\{e_k\}$ of $X_1$ are readily verified to be also eigenvectors
of $L_1$.

On the other hand, we have also another spectrum associated to
(\ref{2.3}). Following \cite{r6,I,IK}, we can write (\ref{2.3}) as
an evolution equation
\begin{equation}
\begin{array}{l}
\ds \dot x(t)=\xi (t)\\
\ds \dot \xi (t)=-\frac{4}{\nu^2}x(t)+\frac{1}{\nu^2}\Big
(X(t,\cos \theta )+X(t,-\cos \theta )\\
\ds +X(t,\sin \theta
)+X(t,-\sin \theta )\Big ) -\frac{1}{\nu^2}f\Big
(\frac{2\pi}{T}t+ x(t)\Big )\\
\ds X_t(t,v)=X_v(t,v)\end{array}\label{5.5}
\end{equation}
on the space
$$
H:=\Big \{(x,\xi,X)\in \R^2\times C^1([-1,1],\R)\mid X(0)=x\Big
\}\, .
$$
The linear part of (\ref{5.5}) is
$$
L=\left (\begin{array}{l}  0 , \qquad 1 ,  \qquad 0 \\ -\frac{4}{\nu^2}  , 0
, \frac{1}{\nu^2}\Big (\partial ^{\cos \theta }+\partial ^{-\cos
\theta }+\partial ^{\sin \theta }+\partial ^{-\sin \theta }\Big )\\
 0 , \qquad 0 , \qquad \partial_v \end{array}\right )\, .
$$
Here $\partial^aX=X(a)$ and $\partial_v$ is the differentiation
with respect to $v$. The spectrum of $L$ is given by the
equation
\begin{equation}
\la^2+\frac{1}{\nu^2}\Big (4 - 2\cosh (\la \cos \theta
)-2\cosh(\la \sin \theta )\Big )=0\, .\label{5.6}
\end{equation}
The imaginary eigenvalues $\la=\imath q$ are determined by
\begin{equation}
\nu^2q^2-4\sin^2\big (\frac{q\cos \theta }{2}\Big )-4\sin^2\big
(\frac{q\sin \theta }{2}\Big )=0\, . \label{5.7}
\end{equation}
Since the nonlinear part of (\ref{5.5}) is $T$-periodic, we see
that condition i) of Theorem \ref{T2.1} are just nonresonance
conditions between the linear and forcing/nonlinear parts of
(\ref{5.5}). This should give an alternative proof of Theorem
\ref{T2.1}.

\

\bibliographystyle{amsplain}

\end{document}